\documentclass{ifacconf}

\usepackage{graphicx}      
\usepackage{natbib}        
\usepackage{amsmath,color} 
\usepackage{amssymb}  
\usepackage{makecell}
\usepackage{booktabs}
\usepackage{fancyhdr}

\fancypagestyle{firstpage}{\lhead{\vspace{-0.75 cm} \small \textbf{{\fontfamily{cmss}\selectfont 10th IFAC Symposium on Robust Control Design (ROCOND)\\August 30-September 2, 2022, Kyoto, Japan}}}}

\begin{document}
\begin{frontmatter}

\title{\large Benefits of Feedforward for Model Predictive Airpath Control of Diesel Engines 
} 


\author[First]{Jiadi Zhang}, 
\author[Second]{Mohammad Reza Amini}, 
\author[First]{Ilya Kolmanovsky},
\author[Third]{Munechika Tsutsumi}, 
\author[Third]{Hayato Nakada}
\vspace{-10pt}
\address[First]{\small Department of Aerospace Engineering, University of Michigan, Ann Arbor, MI 48109, USA \tt(e-mails: {\{jiadi,ilya\}@umich.edu})}
\address[Second]{\small Department of Naval Architecture and Marine Engineering, University of Michigan, Ann Arbor, MI 48109, USA \tt(e-mail: mamini@umich.edu)}
\address[Third]{\small Hino Motors, Ltd., Tokyo 191-8660, Japan \tt({e-mails: \{mu.tsutsumi,hayato.nakada\}@hino.co.jp})}

\begin{abstract}                
This paper investigates options to complement a diesel engine airpath feedback controller with a feedforward. The control objective is to track the intake manifold pressure and exhaust gas recirculation (EGR) rate targets by manipulating the EGR valve and variable geometry turbine (VGT) while satisfying state and input constraints. The feedback controller is based on rate-based Model Predictive Control (MPC) that provides integral action for tracking. Two options for the feedforward are considered one based on a look-up table that specifies the feedforward as a function of engine speed and fuel injection rate, and another one based on a (non-rate-based) MPC that generates dynamic feedforward trajectories. The controllers are designed and verified using a high-fidelity engine model in GT-Power and exploit a low-order rate-based linear parameter-varying (LPV) model for prediction which is identified from transient response data generated by the GT-Power model. It is shown that the combination of feedforward and feedback MPC has the potential to improve the performance and robustness of the control design.  In particular, the feedback MPC without feedforward can lose stability at low engine speeds, while MPC-based feedforward results in the best transient response. Mechanisms by which feedforward is able to assist in stabilization and improve performance are discussed. 
\end{abstract}
\vspace{-4pt}
\begin{keyword}
Model predictive control, Linear parameter-varying control, Diesel airpath control
\end{keyword}

\end{frontmatter}

\section{Introduction}\vspace{-8pt}
In this paper, we consider an airpath control problem for a diesel engine with exhaust gas recirculation (EGR) and variable geometry turbocharging (VGT), and examine complementary characteristics and demonstrate the benefits of combining feedforward and feedback Model Predictive Control (MPC)-based loops.
The feedback MPC loop relies on a rate-based (velocity form) formulation that results in an integral action for offset-free tracking. The feedforward MPC loop exploits the conventional (non-rate-based) MPC formulation.

In particular, we demonstrate that the feedforward MPC that provides the capability of dynamically shaping the input trajectories improves the transient response and tracking performance of the airpath control as compared to a more conventional feedforward implementation using an engine speed-engine fuel injection rate based look-up table.  Furthermore, when operating point changes due to changes in engine speed and fuel injection rate, feedforward MPC helps maintain the trajectory in the region of attraction of the feedback MPC controller and avoid instabilities due to nonlinear behavior of the engine,  thereby improving the robustness of the diesel airpath control. 

Due to its importance for fuel economy improvements and emissions reduction \citep{van2000coordinated}, improvements to the diesel airpath control have been a long-standing research topic in automotive control.  Several MPC solutions have been proposed starting with \cite{ortner2007predictive}. See, for instance, the paper by \cite{huang2018toward} for the historical outlook and 
references.  In particular, the use of switching (non-MPC based) control has been considered in \cite{10.1115/1.1876473} where both feedforward and feedback are switched as the engine is transitioning between different operating points in such a way as to maintain the trajectory in the region of attraction of each operating point. The combined feedforward-feedback MPC architecture for the airpath control has been considered in \cite{liao2020model}, and its investigation is continued in this paper.  

The conclusion about feedforward abetting the stability of the closed-loop system is interesting as regularly feedforward is viewed as a form of an open-loop control and therefore not capable of providing stabilization, with notable exceptions of vibration control \citep{meerkov1980principle}. In constrained control, the use of a sensorless feedforward command governor has been proposed by \cite{garone2011sensorless} that, in principle, could be used to a similar effect, by restricting the trajectory to satisfy constraints informed by the region of attraction dependent on the feedforward. 

\begin{figure}[t!]
\centering
\includegraphics[width=0.45\textwidth]{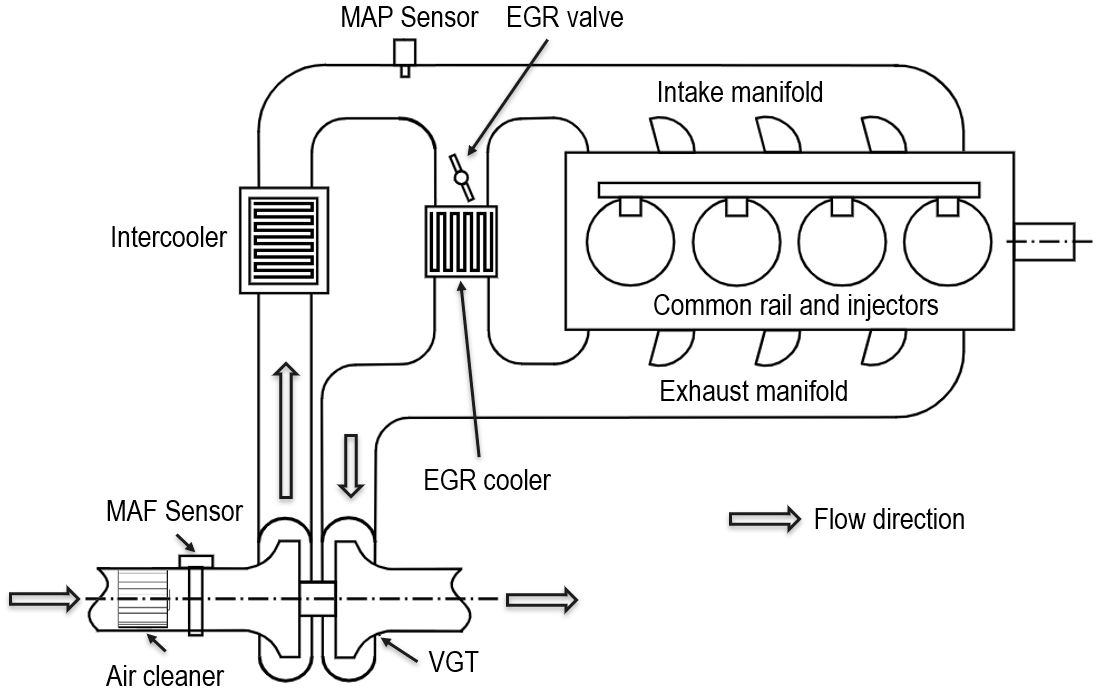}\vspace{-4pt}
\caption{\small Schematic of a 4-cylinder diesel engine with EGR and VGT.}\vspace{-6pt}
\label{fig:engine}
\end{figure}

In this paper, the development of the combined feedforward and feedback MPC solution for the diesel airpath control is performed based on a high fidelity GT-Power engine model which is treated as a black box/surrogate for the engine. The operation-point dependent prediction models are identified based on input-output response data, and the overall model is defined in a linear parameter varying (LPV) form. We purposefully keep the model order low and in the input-output form (states are known outputs) to avoid the need to design and schedule a state observer.  These developments are reported in Section~\ref{sec:sysid}.  The feedforward and feedback MPC design is described in Section~\ref{sec:control}, where we also examine controller tuning issues.  Concluding remarks are made in Section~\ref{sec:conclusion}.  \vspace{-5pt}

\section{Diesel Engine Control Objectives, and Airpath Control-Oriented Modeling} \label{sec:sysid}\vspace{-8pt}
\subsection{Diesel Engine and Airpath Control Objectives}\vspace{-8pt}

Figure~\ref{fig:engine} shows the schematics of a four-cylinder engine considered in this paper. 
The engine consists of a cylinder block, intake and exhaust manifolds, an exhaust gas recirculation (EGR) system, and a variable geometry turbocharger (VGT).  The pressure in the intake manifold and the flow through the compressor of VGT are measured by a manifold absolute pressure (MAP) sensor and a mass airflow (MAF) flow sensor, respectively.

The engine operation point $\rho$ is defined by the engine speed ($N_{\tt e}$) and fuel injection rate ($w_{\tt inj}$) combination, $\rho=[N_{\tt e}, w_{\tt inj}]^{\sf T}$, where $w_{\tt inj}$ corresponds to the sum of the pre-injection and main fuel injection rates. 

The EGR valve controls the mass flow rate from the exhaust manifold into the intake manifold. The VGT controls the intake manifold pressure by varying the exhaust flow area and amount of energy extracted from the exhaust gas. The intake manifold pressure determines the mass flow rate into the engine cylinders. The gas flowing into the engine cylinders consists of air and exhaust gas. 
The EGR rate is defined as
\begin{equation} \label{eq:EGR_rate_def}
\chi_{\tt egr} = {w_{\tt egr}}/{(w_{\tt egr}+w_{\tt c})},
\end{equation}
where $w_{\tt c}$ is the flow rate into the intake manifold through the compressor and intercooler
and $w_{\tt egr}$ is the mass flow through the EGR valve into the intake manifold.  

The objective of the airpath controller is to coordinate the EGR valve and VGT actuators to control the intake manifold pressure and EGR rate to the set-points (targets). These set-points are determined as functions of engine operating point (i.e., engine speed and fuel injection rate) in the engine calibration optimization phase to satisfy emissions and fuel efficiency requirements. Alternatively, these set-points and fuel injection rate could be adjusted by a supervisory controller, such as developed by \cite{liao2020model}, or an economic MPC controller, such as developed by \cite{liao2017cascaded}.
In the sequel, we assume that the intake manifold pressure and EGR rate set-points and the corresponding feedforward EGR valve and VGT positions have been computed and stored in interpolating look-up tables. \vspace{-4pt}

\subsection{Diesel Airpath Control-Oriented Prediction Model}\label{subsec:control-oriented model}\vspace{-4pt}
The MPC implementation requires a control-oriented prediction model.  Following \cite{zhang2022development}, we keep the prediction 
model as simple as possible. Hence the intake manifold pressure ($p_{\tt im}$) and EGR rate ($\chi_{\tt egr}$) are selected as the only model states of the system ($x$).  The model inputs $u$ are EGR valve position (percent open) and VGT position (percent close).  As the intake manifold pressure is measured and the EGR rate is estimated, this model structure eliminates the need for a state observer.
With this approach, the engine constraints need to be remapped as functions of $p_{\tt im}$ and $\chi_{\tt egr}$
which can be done following the approach in \cite{huang2016rate}.

The control-oriented prediction model has an LPV form,\vspace{-6pt}
\begin{eqnarray} \label{eq:model}
x_{k+1} - x_{k+1}^{\tt ss}(\rho_k) & = & A(\rho_k) \left[x_k-x_k^{\tt ss}(\rho_k)\right] \nonumber \\ & + & B(\rho_k) \left[u_k-u_k^{\tt ss}(\rho_k)\right] \nonumber \\ & + & B_f(\rho_k) \left[w_{{\tt inj},k}-w_{{\tt inj}}^{\tt ss}(\rho_k)\right],
\end{eqnarray}
where $k$ denote the discrete time, $\rho_k$ is the vector of engine speed and fuel injection rate at time instant $k$, $A,B: \mathbb{R}^2 \to \mathbb{R}^{2\times2}$,
$x_k^{\tt ss}, u_k^{\tt ss}: \mathbb{R}^2 \to \mathbb{R}^2$ are mappings that determine equilibrium values of $x$ and $u$ corresponding to a given $\rho$,  $B_f:~ \mathbb{R}^2 \to \mathbb{R}^{2 \times 1}$, and  $w_{\tt inj}^{\tt ss}(\rho)=\left[\begin{array}{cc} 0 & 1 \end{array} \right] \rho$.  

The matrices $A(\rho)$ and $B(\rho)$ in (\ref{eq:model}) are computed by linear interpolation of the corresponding matrices of a finite set of $99$ models identified at pre-selected operating points ($\rho$) defined by $9$ values of the engine speed and $11$ values of fuel injection rate that cover the engine operating range.  A high fidelity GT-Power diesel engine model has been used to generate input-output response data corresponding to small input perturbations; these data were then used for local model identification at each of the $99$ operating points. Note that the model (\ref{eq:model}) includes $w_{{\tt inj},k}$ as an extra additive input.  This input has been added as including it improves the model match in transients. For validation results and further details on the LPV airpath model \ref{eq:model} development, see~\cite{zhang2022development}.

\vspace{-4pt}
\section{LPV-based MPC for Airpath Control}\label{sec:control}\vspace{-4pt}
Our MPC design consists of a feedback (\textbf{FB MPC}) loop and a  feedforward. 

\vspace{-5pt}
\subsection{Rate-based Feedback (FB) MPC Loop}
\begin{figure}[h!]
\centering
\includegraphics[width=0.88\columnwidth]{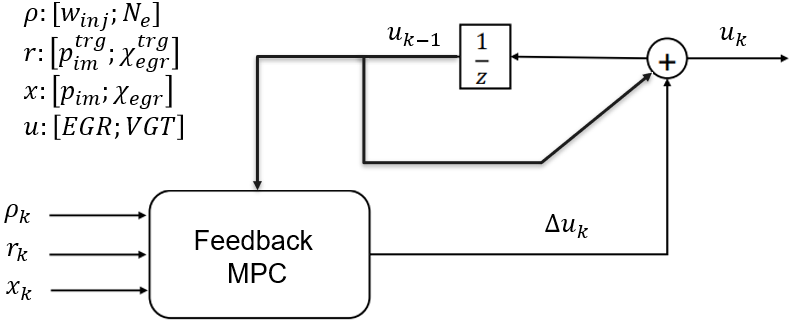}\vspace{-4pt}
\caption{\small Schematic of FB airpath MPC loop. In the square box $z$ stands for the discrete time shift operator, i.e., $u_{k+1} = zu_k$.}\vspace{-4pt}
\label{fig:FB MPC structure}
\end{figure}

\vspace{-4pt}
The FB MPC (Figure~\ref{fig:FB MPC structure}) exploits a rate-based prediction model as in \cite{wang2004tutorial,pannocchia2015offset}. The MPC design based on the rate-based model is able, under appropriate assumptions, to assure offset-free tracking \cite{rawlings2017model} and enhance closed-loop robustness and disturbance rejection capabilities. Such an approach has been also exploited in the earlier work by \cite{huang2016rate}.
\vspace{-4pt}

The basic idea of the rate-based approach is to define rate variables,~$\Delta x_{k} = x_{k} - x_{k-1},~\Delta u_{k} = u_{k} - u_{k-1},$~that correspond to the increments in the state and control variables. Then, assuming $\rho_k$ remains constant over the prediction horizon (it will be updated at the next control update instant), the prediction model (\ref{eq:model}) implies that
\begin{equation} \label{eq:rate base model}
\Delta x_{k+1} = A(\rho_k) \Delta x_k + B(\rho_k) \Delta u_k.
\end{equation}
 Note that  we set $w_{{\tt inj},k}-w_{\tt inj}^{\tt ss}(\rho_k)-
(w_{{\tt inj},k-1}-w_{\tt inj}^{\tt ss}(\rho_{k-1})) \\ = 0$ 
in deriving \eqref{eq:rate base model} from \eqref{eq:model}
as in generating the FB MPC control signal, we assume $\rho_k$ remains constant over the prediction horizon.
The latter assumption is adopted in much of the literature on engine and powertrain control with MPC. See \cite{zhang2022development} for further details.

We let $x_{j|k},$ $u_{j|k}$, $e_{j|k}=x_{j|k}-r_k$, where $r_k$ is the vector of the targets, denote the predicted state, control and tracking error values at time step $j$, $0 \leq j \leq N$, over the prediction horizon when the prediction is made at the time step $k$. Then, to be able to impose constraints on $x_{j|k}$ and $u_{j|k}$ and compute MPC cost that penalizes the predicted tracking error, $e_{j|k}$, we define the augmented state vector,~$\tiny{x_{j|k}^{\tt ext} = \left[\begin{array}{cccc} \Delta x_{j|k}^{\sf T}, & e_{j|k}^{\sf T}, & x_{j-1|k}^{\sf T}, & u_{j-1|k}^{\sf T} \end{array}\right]^{\sf T}}$.~The rate-based prediction model \eqref{eq:rate base model} then implies
\begin{align}
    &x_{j+1|k}^{\tt ext} = \begin{bmatrix}
    A(\rho_k) & 0 & 0 & 0\\
    A(\rho_k) & \mathbb{I}_{n_e\times n_e} & 0 & 0\\
    \mathbb{I}_{n_x\times n_x} & 0 & \mathbb{I}_{n_x\times n_x} & 0\\
    0 & 0 & 0 & \mathbb{I}_{n_u\times n_u}
    \end{bmatrix}
    x_{j|k}^{\tt ext}  \nonumber\\ 
    & \qquad \qquad + \begin{bmatrix}
    B(\rho_k) \\
    B(\rho_k) \\
    0\\
    \mathbb{I}_{n_u\times n_u}
    \end{bmatrix} \Delta u_{j|k},
\end{align}
where $\Delta u_{j|k}$ is the control input in the extended system.

The FB MPC design is based on the solution of the following discrete time optimal control problem:
\begin{subequations}\label{eq:modified MPC}
\begin{multline}
    \min_{\Delta u_{0|k},...,\Delta u_{N-1|k},\epsilon_k} (x_{N|k}^{\tt ext})^{\sf T} P_{\infty|k}x_{N|k}^{\tt ext} +  \\
    \sum_{j=0}^{N-1} (x_{j|k}^{\tt ext})^{\sf T} Q^{\tt ext}x_{j|k}^{\tt ext} + \Delta u_{j|k}^{\sf T} R^{\tt ext}\Delta u_{j|k} + \mu\epsilon_k^{\sf T} \epsilon_k
\tag{\ref{eq:modified MPC}}
\end{multline}

subject to
\begin{align} 
    &\Delta x_{0|k} = x_{k} - x_{k-1} \\
    &e_{0|k} = x_{k} - r_k\\
    &x_{-1|k} = x_{k-1}\\
    &u_{-1|k} = u_{k-1} \label{subeq:u}\\  
    &\Delta x_{j+1|k} = A(\rho_k)\Delta x_{j|k} + B(\rho_k)\Delta u_{j|k}\\
    &e_{j+1|k} = A(\rho_k)\Delta x_{j|k} + B(\rho_k)\Delta u_{j|k} + e_{j|k}\\
    &x_{j|k} = x_{j-1|k} + \Delta x_{j|k}, j = 1, \ldots, N,\\
    &u_{j|k} = u_{j-1|k} + \Delta u_{j|k}, j = {0}, \ldots, N-1,\\ 
    &x_{min} - \epsilon_k\leq x_{j|k} \leq x_{max} + \epsilon_k, j = 1, \ldots, N, \label{equ:cnrsrate}\\
    &u_{min} \leq u_{j|k} \leq u_{max}, j = 0,\ldots, N-1,
\end{align}
\end{subequations}

where $N$ is the prediction horizon,  $Q^{\tt ext} \succeq 0$ and $R^{\tt ext} \succ 0$ are state and control weighting matrices, with $Q^{\tt ext}$ chosen so that~$\tiny{(x_{j|k}^{\tt ext})^{\sf T} Q^{\tt ext} x_{j|k}^{\tt ext} = e_{j|k}^{\sf T} Q_e e_{j|k}}$,~and $\epsilon_k$ is the slack variable introduced to avoid infeasibility of the state constraints. 

The matrix $P_{\infty|k} \succeq 0$ imposes the terminal penalty on $\Delta x_{N|k}$ and $e_{N|k}$ states; such a terminal penalty is beneficial to ensure local closed-loop stability (when constraints are inactive with the engine operating near the set-point). To compute $P_{\infty|k}$, we first compute a matrix $\tilde{P}_{\infty|k}$ by solving the discrete algebraic Riccati equation (DARE) corresponding to the model for the evolution of just $\Delta x_{j|k}$ and $e_{j|k}$ states for which the dynamics ($A$) and input ($B$) matrices have the
form,
\begin{equation}
\begin{bmatrix}
A(\rho_k) & 0\\
A(\rho_k) & \mathbb{I}_{n_e\times n_e}
\end{bmatrix}
, 
\begin{bmatrix}
B(\rho_k)\\
B(\rho_k)
\end{bmatrix},
\end{equation}
while the state and control weighting matrices are given by\vspace{-4pt}
\begin{equation}
\begin{bmatrix}
0 & 0\\
0 & Q_e
\end{bmatrix}
, 
R^{\tt ext}.
\end{equation}
{The terminal penalty matrix can be computed for a discrete set of values of $\rho$ on a chosen mesh and then interpolated element by element.}  

Numerically, \eqref{eq:modified MPC} reduces to a quadratic programming (QP) problem.  The first element of the optimal control sequence, $\Delta u_{0|k}^*$ then informs the FB MPC control signal according to the relation,~$u_k=u_{k-1} + \Delta u_{0|k}^*$.

\begin{figure}[h!]
\centering
\includegraphics[width=0.44\textwidth]{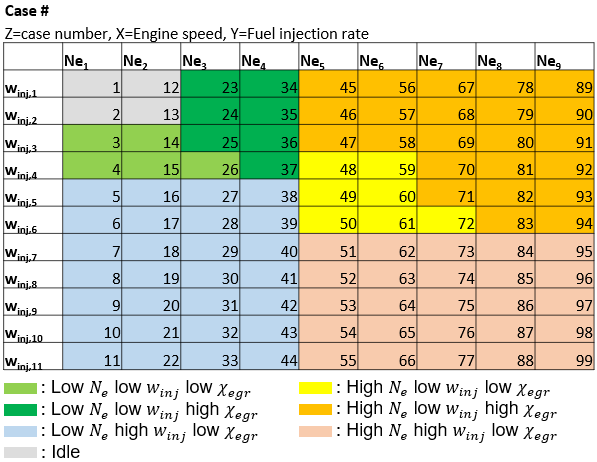}\vspace{-5pt}
\caption{\small Seven regions defined as a function of  $N_{\tt e}$, $w_{\tt inj}$ and $\chi_{\tt egr}$, to simplify the MPC tuning process with the same $Q^{ext}$ and $R^{ext}$ pair used in each region.}\vspace{-8pt}
\label{fig:QR}
\end{figure}

While the changes in the dynamic characteristics of the engine response across the engine operating range are captured by our LPV prediction model, we have also found in~\cite{zhang2022development} that the closed-loop performance can be further improved by varying
the weights { $Q_{e}$} and $R^{\tt ext}$ in \eqref{eq:modified MPC}. To make the tuning process more tractable, we chose to group multiple operating points into regions and then tune the weighting matrices with the restriction that $Q_e$ (i.e., $Q^{\tt ext}$) and $R^{\tt ext}$ are the same for all operating points in a given region. In our implementation, we defined seven regions as a function of  $N_{\tt e}$, $w_{\tt inj}$ and $\chi_{\tt egr}$, as shown in Figure~\ref{fig:QR}. Note that the values of  $N_{\tt e}$, $w_{\tt inj}$ and $\chi_{\tt egr}$ are categorized using qualitative ``high'' and ``low'' terms to protect the OEM proprietary data.
\vspace{-8pt}

\begin{figure}[h!]
\centering
\includegraphics[width=0.44\textwidth]{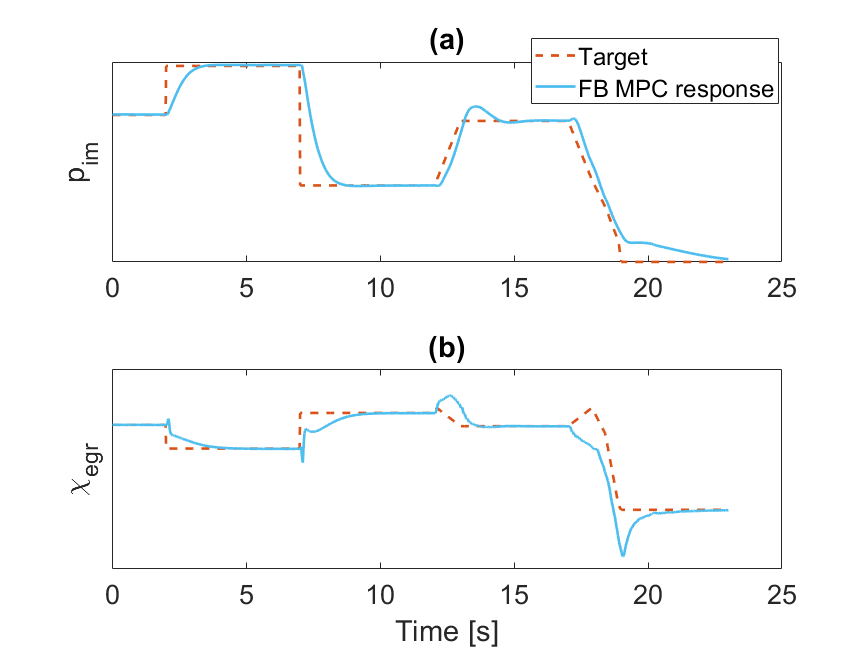}\vspace{-5pt}
\caption{\small Example of (a) $p_{\tt im}$ and (b) $\chi_{\tt egr}$ trajectory tracking with FB MPC in response to steps in $w_{\tt inj}$ {followed by ramps in $N_{\tt e}$}.}\vspace{-3pt}
\label{fig:simple case study one result}
\end{figure}

Figure~\ref{fig:simple case study one result} illustrates closed-loop response to steps in $w_{\tt inj}$ {followed by ramps in $N_{\tt e}$}  with the fully tuned FB MPC { co-simulated with the GT-Power engine model through the Simulink interface}. In these and subsequent simulations, we use the sampling period of $20$ ms and prediction horizon of $1$ s.
The GT-Power simulation is configured to provide cycle averaged signals.  The package {\tt mpctools} developed by \cite{risbeck2016mpctools} is used for the numerical solution of MPC problem.
The set-points are tracked with zero steady-state error and the 
overshoot and settling time of $p_{\tt im}$ are within the acceptable range ($1.5$ s).  However, the response of  $\chi_{\tt egr}$ to a step-change in $w_{\tt inj}$ has an undershoot and the tracking error is large in transients.  The transient response can be made faster with a more aggressive tuning, however, in this case, it starts exhibiting larger overshoot and oscillations as is typical of controllers with integral action.

While the capability to improve closed-loop transient response by tuning of FB MPC is limited, there is an alternative route which is to exploit feedforward. This is further considered in the next section.

\vspace{-3pt}

\subsection{Combination of Feedforward Look-up Table and FB MPC }\vspace{-5pt}
Frequently, the engine control strategy involves a combination of feedforward and feedback. A feedback controller with an integral action alone is capable of eliminating the steady-state error, but its response can be slow.~
The combination of the feedforward and feedback can produce a faster action by the actuators and improve transient response. 

In the case of our diesel engine, the simplest feedforward consists of steady-state EGR valve and VGT positions, consistent with $p_{\tt im}$ and $\chi_{\tt egr}$ set-points (and with given engine speed and fuel injection rate values) in steady-state, and aggregated in a look-up table that computes them as a function of engine speed and fuel injection rate through the linear interpolation. We refer to such a feedforward (FF) solution as a ``look-up table FF''.
 
\begin{figure}[h!]
        \centering
        \includegraphics[width=0.92\columnwidth]{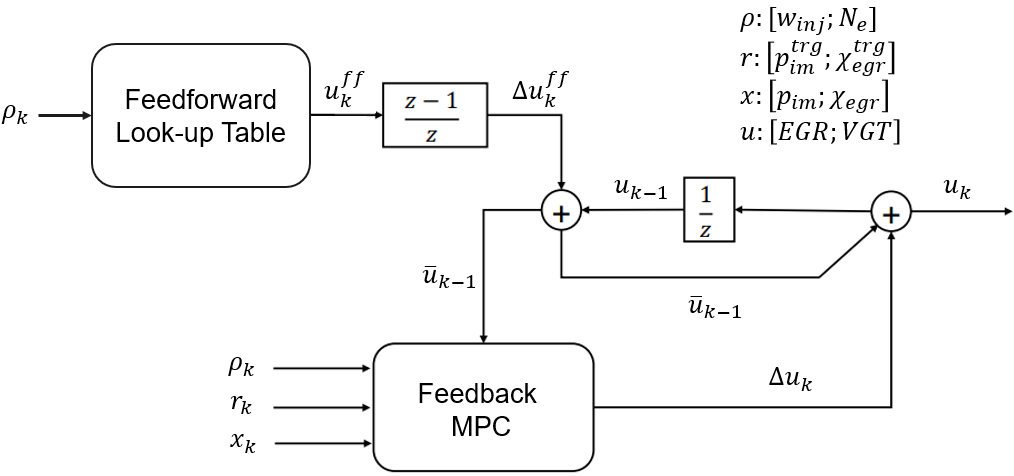}\vspace{-5pt}
        \caption{\small Schematic of look-up table FF + FB MPC architecture. Note that $\Delta u_k=\Delta u_{0|k}$ is computed by FB MPC, $u^{\tt ff}_{k}$ is  generated by the FF look-up table, $u_k$ is applied to the engine, and $\bar{u}_{k-1}$ 
        defined by \eqref{eq:modified u} is provided to FB MPC.}\vspace{-3pt}
        \label{fig:FF look-up with FB MPC structure}
\end{figure}

The integration of FB MPC with look-up table FF is achieved as shown in 
Figure~\ref{fig:FF look-up with FB MPC structure}.  The modification of the rate-based FB MPC occurs in the expression \eqref{subeq:u} which is replaced by
       \begin{equation}\label{eq:modified u}
           u_{-1|k} = \bar{u}_{k-1} = u_{k-1} + u_k^{\tt ff} - u_{k-1}^{\tt ff},
       \end{equation}
       where $u_k^{\tt ff}$ is the output of the FF controller at the time instant $k$. The control signal at the time instant $k$ is computed according to
\begin{equation}\label{eq:mod1}
 u_{k} = \bar{u}_{k-1} + \Delta u_{0|k}^*. 
 \end{equation}
\vspace{-7pt}
        
\begin{figure}[h!]
    \centering
    \includegraphics[width=0.44\textwidth]{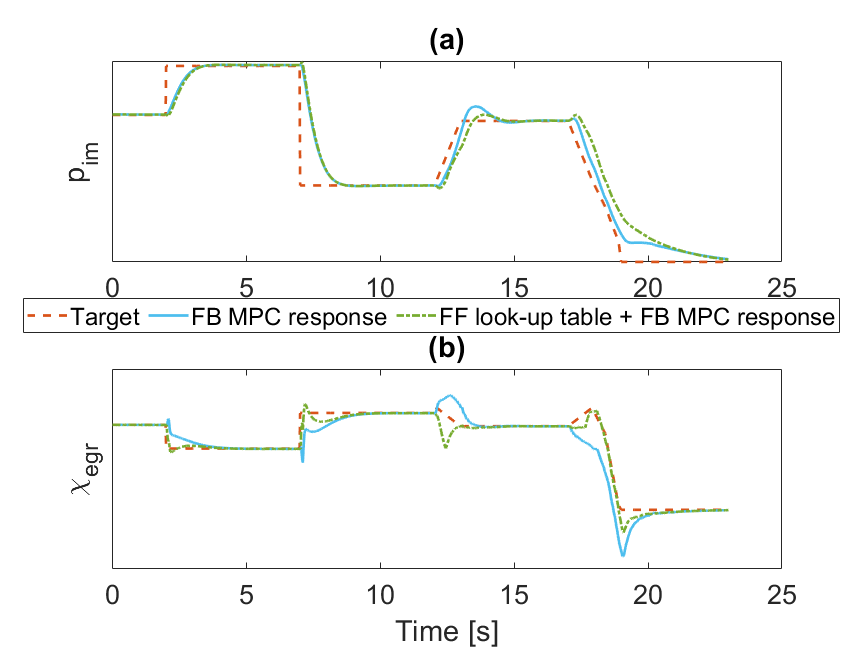}\vspace{-7pt}
    \caption{\small Example of (a) $p_{\tt im}$ and (b) $\chi_{\tt egr}$ trajectory tracking with look-up table FF and FB MPC during {first, a fuel ($w_{\tt inj}$) tip-in and tip-out and then, a ramp change of $N_{\tt e}$.}}\vspace{-4pt}
    \label{fig:simple case study two results}
    \end{figure}

As shown in Figure~\ref{fig:simple case study two results} for the same simple scenario as in Figure~\ref{fig:simple case study one result} {and the same tuning of FB MPC}, the look-up table FF was able to speed up $\chi_{\tt egr}$ tracking during fuel tip-in and tip-out and has greatly reduced the tracking error of $\chi_{\tt egr}$ during the fast {ramp change of $N_{\tt e}$.}  For $p_{\tt im}$ tracking, the look-up table FF reduced the overshoot during the increase of $N_{\tt e}$ but it increased the tracking error during the decrease of $N_{\tt e}$.  More complex simulation scenarios will be treated in Section \ref{sec:results}.
\vspace{-7pt}
    
We note that in the absence of the model mismatch, the conventional MPC guarantees that 
the true next state and predicted one-step ahead states are the same, i.e., 
$x_{k+1}=x_{1|k}$.  This property is not maintained by the considered implementation if  $u_k^{\tt ff} \neq u_{k-1}^{\tt ff}$. More specifically, it can be easily shown that the true state at time instant $k+1$ is given by
\begin{align}\label{equ:ffwd_effect}
x_{k+1}=x_{1,k}+B(\rho_k)(u_k^{\tt ff}-u_{k-1}^{\tt ff}),\\
x_{1|k}=A(\rho_k)x_k+B(\rho_k)u_{k-1}+B\Delta u_{0|k}^*,
\end{align} 
even though $u_{k}=u_{0|k}=u_{k-1}+\Delta u_{0|k}^*+u_k^{\tt ff}-u_{k-1}^{\tt ff}$.  The equation (\ref{equ:ffwd_effect}) clarifies that the feedforward is able to affect the state of the system.  From the perspective of the rate-based MPC, the term $(u_k^{\tt ff}-u_{k-1}^{\tt ff})$ appears as a disturbance; to accommodate it, under the assumption that $u_{k+j}^{\tt ff}=u_{k+j-1}^{\tt ff}$ over the prediction horizon (consistent with the assumption of the operating point not changing), the state constraints (\ref{equ:cnrsrate}) could be imposed on the prediction, $\bar{x}_{j|k}$, of the true state, given by
$$\bar{x}_{j|k}=x_{j|k}+\left[\sum_{m=0}^{j-1} A(\rho_k)^{j-1-m} B(\rho_k)\right](u_k^{\tt ff}-u_{k-1}^{\tt ff}).$$
This in effect tightens the constraints to accommodate changes in the feedforward. In the subsequent simulations, we did not pursue this constraint tightening approach to avoid conservatism and enforced the constraints (\ref{equ:cnrsrate}).
  
\begin{figure}[h!]
    \centering
    \includegraphics[width=0.44\textwidth]{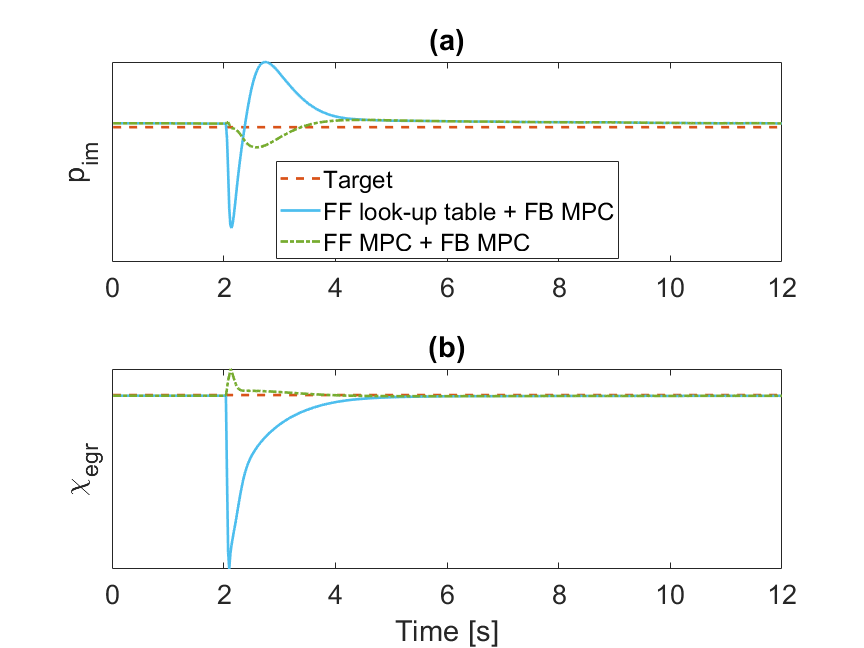}\vspace{-6pt}
    \caption{\small Example of (a) $p_{\tt im}$ and (b) $\chi_{\tt egr}$ trajectory tracking with FF look-up table and FB MPC during a step change of $w_{\tt inj}$ and constant $N_{\tt e}$. A different set of target $p_{\tt im}$ and $\chi_{egr}$ is used here without re-calibrating the FF look-up table.}\vspace{-4pt}
    \label{fig:change of target}
    \end{figure}

The solution based on look-up table FF for EGR valve and VGT positions has a drawback in that if at a given operating point $\rho_k$, the targets $r_k$ need to be different from the one assumed  (e.g., EGR rate needs to be increased to satisfy different emissions regulations or during engine warm-up), the look-up table FF will no longer be accurate.  The look-up table used for FF may also become inaccurate due to part-to-part variability or aging.  To illustrate potential issues this may cause, Figure~\ref{fig:change of target}
provides simulation results for the case of constant $N_{\tt e}$ and a step-change in $w_{\tt inj}$ when the set-points for $p_{\tt im}$ and $\chi_{\tt egr}$ do not change to follow the change in $w_{\tt inj}$.  The responses of the combined look-up table FF and FB MPC now exhibit large and undesirable overshoot.  As this example illustrates, the mismatch of feedforwarded EGR valve and VGT positions and intake manifold pressure and EGR rate set-points  
can degrade the closed-loop transient response.  The need to eliminate such mismatches increases calibration time and effort and can potentially necessitate a more complex implementation of the look-up table FF (e.g., multi-dimensional or adaptive look-up table).  In the next section, we consider a different approach to implementing feedforward that addresses the above issue.
\vspace{-6pt}

\subsection{Combination of FF MPC and FB MPC }
\begin{figure}[h!]
        \centering
        \includegraphics[width=0.9\columnwidth]{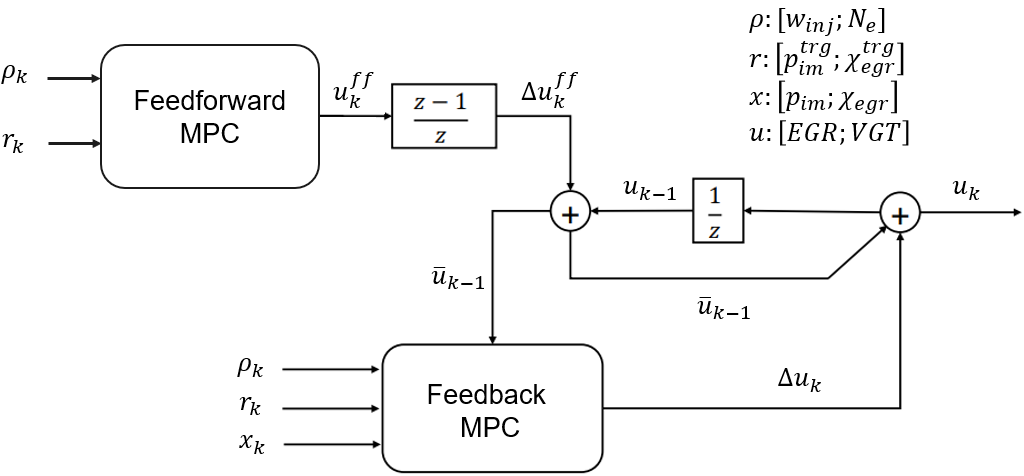}\vspace{-6pt}
        \caption{\small Schematic of FF MPC  + FB MPC architecture. Note that $\Delta u$ is computed from the FB MPC, {$u^{\tt ff}$ is generated by FF MPC}, $u$ is applied to the engine (GT-Power model), and $\bar{u}$ is supplied to the FB controller through \eqref{eq:modified u}.}\vspace{-4pt}
        \label{fig:FF MPC with FB MPC structure}
\end{figure}

\vspace{-6pt}
The feedforward MPC (\textbf{FF MPC}) and FB MPC combination is illustrated in Figure~\ref{fig:FF MPC with FB MPC structure}.  Unlike FB MPC, the FF MPC only closes the loop around the model of the plant, i.e., it does not use the actual engine measured signals as inputs.  The primary objective of FF MPC is to approximately shape the input trajectory as needed to evolve the trajectory towards the targets.

The FF MPC is using the discrete-time model (\ref{eq:model}) for prediction. The control action is defined by solving the following
discrete-time optimal control problem:\vspace{-4pt}
    \begin{subequations}\label{eq:origional MPC}
    \begin{multline} 
    \min_{\tilde{u}_{0|k},...,\tilde{u}_{N-1|k}} e_{N|k}^{\sf T} P^{\tt ff}_{\infty|k}e_{N|k} + \sum_{j=0}^{N-1}  e_{j|k}^{\sf T} Q^{\tt ff} e_{j|k} + \\ \tilde{u}_{j|k}^{\sf T} R^{\tt ff} \tilde{u}_{j|k}
    \tag{\ref{eq:origional MPC}}
    \end{multline}
    subject to \vspace{-4pt}
    \begin{align} 
    &e_{j|k} = \tilde{x}_{j|k} - \tilde{r}_{j|k}\\
    &\tilde{x}_{j|k} = x_{j|k} - x_k^{\tt ss}\\
    &\tilde{r}_{j|k} = r_{j|k} - x_k^{\tt ss}\\
    & {r_{j|k}=r_k} \\
    &\tilde{u}_{j|k} = u_{j|k} - u_k^{\tt ss}\\
    &e_{0|k} = \tilde{x}_{k} - \tilde{r}_k\\
    &e_{j+1|k} = A(\rho_k) \tilde{x}_{j|k} + B(\rho_k)\tilde{u}_{j|k} + e_{j|k}\\
    &x_{\tt min} \leq x_{j|k} \leq x_{\tt max}, j = 1,..., N,\\
    &u_{\tt min} \leq u_{j|k} \leq u_{\tt max}, j = 0,..., N-1.
    \end{align}
    \end{subequations}
{The final control signal is determined by (\ref{eq:mod1}) and 
(\ref{eq:modified MPC}) with $u^{\tt ff}_k=u^*_{0|k}$, where
$u^*_{0|k}$ is the first element of the optimal solution sequence
to (\ref{eq:origional MPC}).
}

\begin{figure}[h!]
    \centering
    \includegraphics[width=0.44\textwidth]{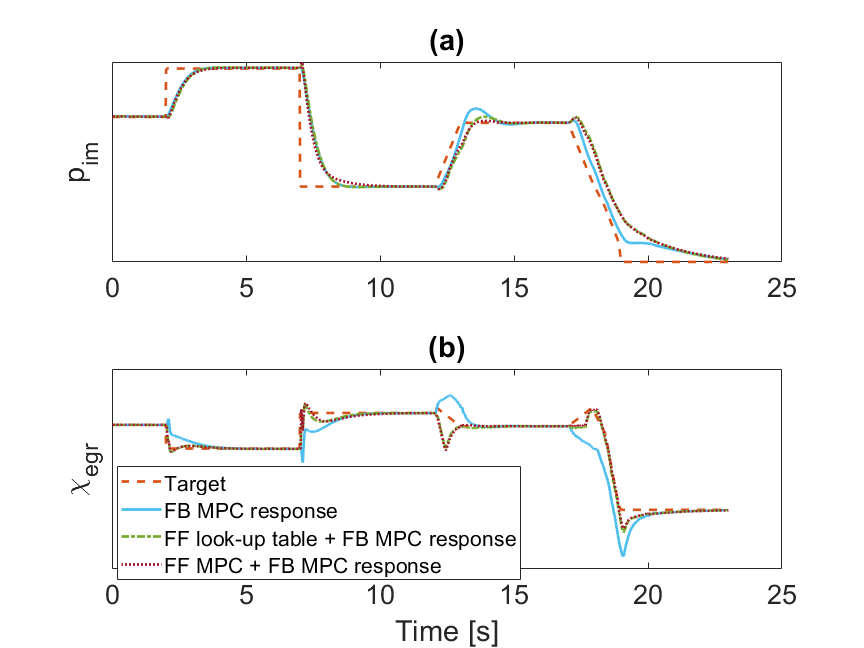} \vspace{-5pt}
    \caption{\small Example of (a) $p_{\tt im}$ and (b) $\chi_{\tt egr}$ trajectory tracking with MPC FF and FB MPC during {first, a fuel ($w_{\tt inj}$) tip-in and tip-out and then, a ramp change of $N_{\tt e}$.}}\vspace{-5pt}
    \label{fig:simple case study three result}
    \end{figure}

Simulation results of the closed-loop system with FF MPC and FB MPC combination are shown in Figure~\ref{fig:simple case study three result} for the same simple scenario as in Figure~\ref{fig:simple case study one result}
and Figure~\ref{fig:simple case study two results}
{and same tuning of FB MPC.}  {
The responses with FF MPC match those with look-up table FF and are improved as compared to only FB MPC. More complex simulation scenarios will be treated in Section \ref{sec:results}.}
    \vspace{-9pt}

We note that more advanced options for integrating FF MPC and FB MPC may exist, e.g., using a sequence generated by FF MPC to warm-start FB MPC.
We leave the investigation of such options to future work.

\vspace{-6pt}
\section{Simulation Results} \label{sec:results}
\begin{figure}[h!]
\centering
\includegraphics[width=0.95\columnwidth]{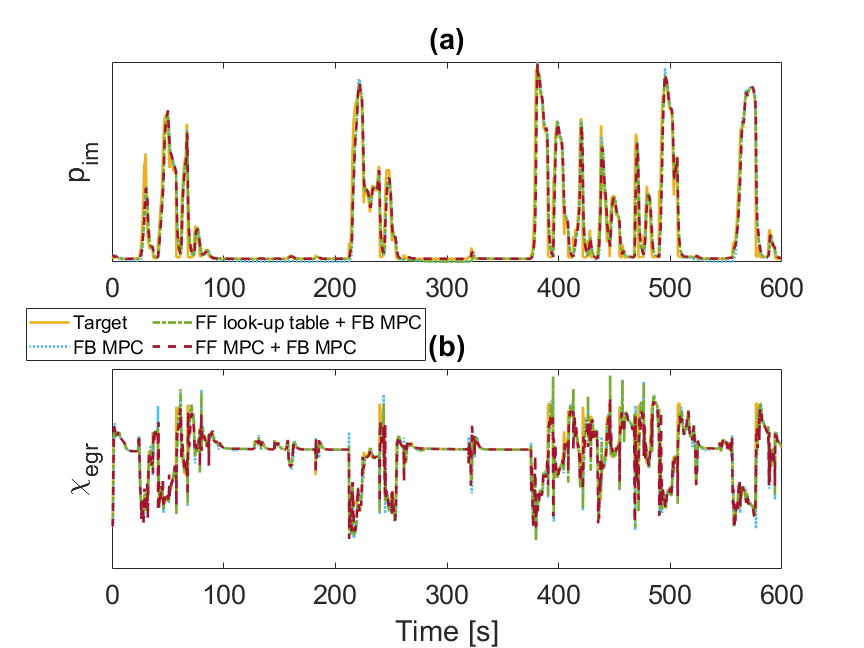}\vspace{-6pt}
\caption{\small Time histories of (a) $p_{\tt im}$ and (b) $\chi_{\tt egr}$ with three different control schemes over the first 600 s of FTP simulation.}\vspace{-8pt}
\label{fig:FTP result}
\end{figure}

\begin{figure}[h!]
\centering
\includegraphics[width=0.95\columnwidth]{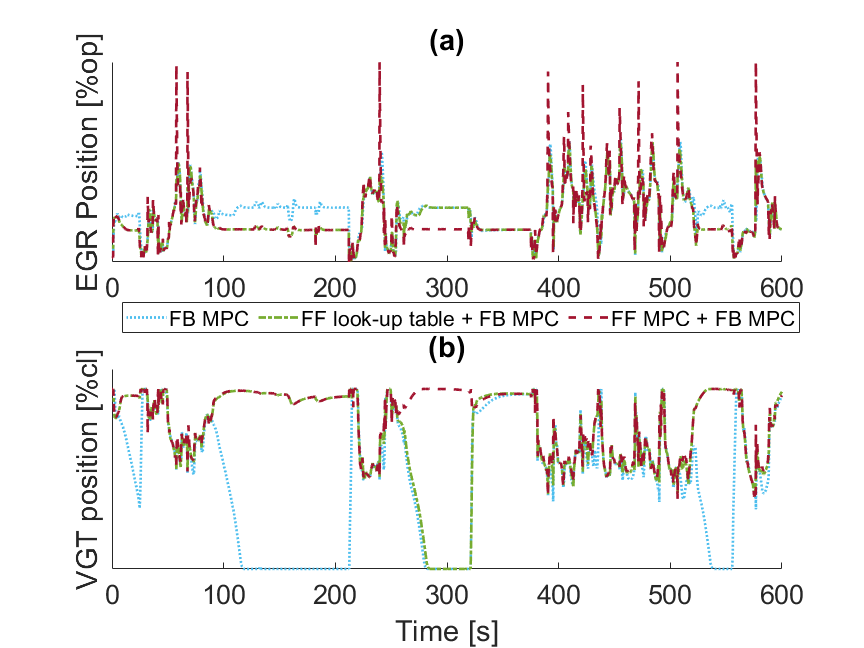}\vspace{-6pt}
\caption{\small Time histories of (a) $EGR$ and (b) $VGT$ position with three different control schemes as in  Figure~\ref{fig:FTP result} over the first 600 s of FTP simulation.}\vspace{-3pt}
\label{fig:FTP input}
\end{figure}

\vspace{-6pt}
We now evaluate the three control designs (FB MPC only, look-up table FF and FB MPC, and FF MPC and FB MPC) in more complex simulation scenarios over Federal Test Procedure (FTP)~
drive cycle. The results are summarized in
Figures~\ref{fig:FTP result}-\ref{fig:FTP input} and Table~\ref{tbl:FTP}.
In terms of {average absolute tracking errors}, the best tracking performance is
achieved by the controller that combines FF MPC and FB MPC.

Of particular interest are large VGT excursions ending in VGT being fully open that are observed near idle with FB MPC only controller and to a lesser degree with look-up table FF and FB MPC combination. While set-point tracking is approximately maintained, these large deviations in the VGT position are indicative of potential closed-loop instability, likely caused by the state trajectory leaving the region of attraction provided by FB MPC around the current operating point $\rho_k$ when $\rho_k$ undergoes rapid transients. At the same time, FF MPC is able to to evolve the trajectory sufficiently towards new operating point to avoid instability.
\vspace{-2pt}  
\vspace{-8pt}  

\begin{table}[h!]
\caption{\small Comparison of three control schemes over the FTP.}\vspace{-5pt}
\label{tbl:FTP}
{\small
\begin{center}
\begin{tabular}{lll}
\toprule  
\makecell[l]{\textbf{Controller}} & \makecell[l]{\textbf{ $\overline{e}_{p_{\tt im}}$}[bar]} & \makecell[l]{\textbf{$\overline{e}_{\chi_{\tt egr}}$}} \\
\midrule  
\textbf{FB MPC Only} & 0.0771 & 0.0144\\ 
(reference) &  & \\ \hline
\textbf{Look-up table FF} & 0.0692 & 0.0117\\
\textbf{+ FB MPC} & ($\downarrow$ 10.3\%) & ($\downarrow$ 18.8\%)\\\hline
\textbf{FF MPC} & 0.0650 & 0.0103\\
\textbf{+ FB MPC} & ($\downarrow$ 15.7\%) & ($\downarrow$ 28.5\%)\\
\bottomrule 
\end{tabular}\vspace{-1pt}
\end{center}}
\end{table}  

\section{Conclusions}\label{sec:conclusion}\vspace{-5pt}
Complimenting the feedback controller based on rate-based Model Predictive Control (MPC)  with the feedforward has the potential to improve the tracking performance of diesel engine airpath control systems.
Such a feedforward can be based either on a look-up table or 
a (non-rate-based)  feedforward MPC.  The latter option generates dynamic feedforward trajectories and is more effective in reducing tracking errors while eliminating potential closed-loop instabilities at low engine speed by helping the state back to the region of attraction of the feedback controller.

\begin{ack}\vspace{-5pt}
The authors would like to acknowledge useful 
discussions with Dr. Dominic Liao-McPherson
on the topic of diesel engine control using MPC.
\end{ack}

\vspace{-5pt}
\bibliography{ifacconf}            

\begin{thebibliography}{14}
\providecommand{\natexlab}[1]{#1}
\providecommand{\url}[1]{\texttt{#1}}
\providecommand{\urlprefix}{URL }
\expandafter\ifx\csname urlstyle\endcsname\relax
  \providecommand{\doi}[1]{doi:\discretionary{}{}{}#1}\else
  \providecommand{\doi}{doi:\discretionary{}{}{}\begingroup
  \urlstyle{rm}\Url}\fi

\bibitem[{Bengea et~al.(2004)Bengea, DeCarlo, Corless, and
  Rizzoni}]{10.1115/1.1876473}
Bengea, S., DeCarlo, R., Corless, M., and Rizzoni, G. (2004).
\newblock {A Polytopic system approach for the hybrid control of a diesel
  engine using VGT/EGR}.
\newblock \emph{Journal of Dynamic Systems, Measurement, and Control}, 127(1),
  13--21.

\bibitem[{Garone et~al.(2011)Garone, Tedesco, and
  Casavola}]{garone2011sensorless}
Garone, E., Tedesco, F., and Casavola, A. (2011).
\newblock Sensorless supervision of linear dynamical systems: The feed-forward
  command governor approach.
\newblock \emph{Automatica}, 47(7), 1294--1303.

\bibitem[{Huang et~al.(2018)Huang, Liao-McPherson, Kim, Butts, and
  Kolmanovsky}]{huang2018toward}
Huang, M., Liao-McPherson, D., Kim, S., Butts, K., and Kolmanovsky, I. (2018).
\newblock Toward real-time automotive model predictive control: A perspective
  from a diesel air path control development.
\newblock In \emph{American Control Conference}.
\newblock Milwaukee, WI, USA.

\bibitem[{Huang et~al.(2016)Huang, Zaseck, Butts, and
  Kolmanovsky}]{huang2016rate}
Huang, M., Zaseck, K., Butts, K., and Kolmanovsky, I. (2016).
\newblock Rate-based model predictive controller for diesel engine air path:
  Design and experimental evaluation.
\newblock \emph{IEEE Transactions on Control Systems Technology}, 24(6),
  1922--1935.

\bibitem[{Liao-McPherson et~al.(2020)Liao-McPherson, Huang, Kim, Shimada,
  Butts, and Kolmanovsky}]{liao2020model}
Liao-McPherson, D., Huang, M., Kim, S., Shimada, M., Butts, K., and
  Kolmanovsky, I. (2020).
\newblock Model predictive emissions control of a diesel engine airpath: Design
  and experimental evaluation.
\newblock \emph{International Journal of Robust and Nonlinear Control}, 30(17),
  7446--7477.

\bibitem[{Liao-McPherson et~al.(2017)Liao-McPherson, Kim, Butts, and
  Kolmanovsky}]{liao2017cascaded}
Liao-McPherson, D., Kim, S., Butts, K., and Kolmanovsky, I. (2017).
\newblock A cascaded economic model predictive control strategy for a diesel
  engine using a non-uniform prediction horizon discretization.
\newblock In \emph{2017 IEEE Conference on Control Technology and Applications
  (CCTA)}, 979--986. IEEE.

\bibitem[{Meerkov(1980)}]{meerkov1980principle}
Meerkov, S. (1980).
\newblock Principle of vibrational control: theory and applications.
\newblock \emph{IEEE Transactions on Automatic Control}, 25(4), 755--762.

\bibitem[{Ortner and Del~Re(2007)}]{ortner2007predictive}
Ortner, P. and Del~Re, L. (2007).
\newblock Predictive control of a diesel engine air path.
\newblock \emph{IEEE transactions on control systems technology}, 15(3),
  449--456.

\bibitem[{Pannocchia et~al.(2015)Pannocchia, Gabiccini, and
  Artoni}]{pannocchia2015offset}
Pannocchia, G., Gabiccini, M., and Artoni, A. (2015).
\newblock Offset-free mpc explained: novelties, subtleties, and applications.
\newblock \emph{IFAC-PapersOnLine}, 48(23), 342--351.

\bibitem[{Rawlings et~al.(2017)Rawlings, Mayne, and Diehl}]{rawlings2017model}
Rawlings, J.B., Mayne, D.Q., and Diehl, M. (2017).
\newblock \emph{Model predictive control: theory, computation, and design},
  volume~2.
\newblock Nob Hill Publishing Madison, WI.

\bibitem[{Risbeck and Rawlings(2016)}]{risbeck2016mpctools}
Risbeck, M. and Rawlings, J. (2016).
\newblock {MPCTools: Nonlinear Model Predictive Control Tools for CasADi}.
\newblock {[online] Available:
  \tt\urlstyle{bitbucket.org/rawlings-group/octave-mpctools}}.

\bibitem[{Van~Nieuwstadt et~al.(2000)Van~Nieuwstadt, Kolmanovsky, and
  Moraal}]{van2000coordinated}
Van~Nieuwstadt, M.J., Kolmanovsky, I.V., and Moraal, P.E. (2000).
\newblock Coordinated egr-vgt control for diesel engines: an experimental
  comparison.
\newblock \emph{SAE Transactions}, 238--249.

\bibitem[{Wang(2004)}]{wang2004tutorial}
Wang, L. (2004).
\newblock A tutorial on model predictive control: Using a linear velocity-form
  model.
\newblock \emph{Developments in Chemical Engineering and Mineral Processing},
  12(5-6), 573--614.

\bibitem[{Zhang et~al.(2022)Zhang, Amini, Kolmanovsky, Tsutsumi, and
  Nakada}]{zhang2022development}
Zhang, J., Amini, M.R., Kolmanovsky, I., Tsutsumi, M., and Nakada, H. (2022).
\newblock Development of a model predictive airpath controller for a diesel
  engine on a high-fidelity engine model with transient thermal dynamics.
\newblock In \emph{American Control Conference}.
\newblock Atlanta, GA, USA.

\end{thebibliography}

\end{document}